\documentclass[conference]{IEEEtran}
\IEEEoverridecommandlockouts
\usepackage{amsmath,amssymb,amsfonts}
\usepackage{algorithmic}
\usepackage{graphicx}
\usepackage{textcomp}
\usepackage{xcolor}
\usepackage{multirow}
\usepackage{svg}
\usepackage{enumitem}
\usepackage{caption}
\usepackage{booktabs}
\usepackage{siunitx}
\usepackage{authblk}
\usepackage[numbers,sort&compress]{natbib}

\def\BibTeX{{\rm B\kern-.05em{\sc i\kern-.025em b}\kern-.08em
    T\kern-.1667em\lower.7ex\hbox{E}\kern-.125emX}}
\begin{document}
\renewcommand\Authands{, } 
\title{Reference Channel Selection by Multi-Channel Masking for End-to-End Multi-Channel Speech Enhancement
}

\author[1]{Wang Dai}
\author[2]{Xiaofei Li}
\author[1]{Archontis Politis}
\author[1]{Tuomas Virtanen}
\affil[1]{Audio Research Group, Tampere University, Tampere, Finland}
\affil[2]{Westlake University \& Westlake Institute for Advanced Study, Hangzhou, China}
\affil[ ]{\texttt{\{name.surname\}@tuni.fi}, \texttt{lixiaofei@westlake.edu.cn}}

\maketitle

\begin{abstract}
In end-to-end multi-channel speech enhancement, the traditional approach of designating one microphone signal as the reference for processing may not always yield optimal results. The limitation is particularly in scenarios with large distributed microphone arrays with varying speaker-to-microphone distances or compact, highly directional microphone arrays where speaker or microphone positions change over time. Current mask-based methods often fix the reference channel during training, which makes it not possible to adaptively select the reference channel for optimal performance. To address this problem, we introduce an adaptive approach for selecting the optimal reference channel.
Our method leverages a multi-channel masking-based scheme, where multiple masked signals are combined to generate a single-channel output signal. This enhanced signal is then used for loss calculation, while the reference clean speech is adjusted based on the highest scale-invariant signal-to-distortion ratio (SI-SDR).
The experimental results on the Spear challenge simulated dataset D4 demonstrate the superiority of our proposed method over the conventional approach of using a fixed reference channel with single-channel masking.
\end{abstract}


\section{Introduction}
Multi-channel speech enhancement leverages information from multiple microphones to enhance the target speech signal while suppressing noise, reverberation and interference. Common applications include speech recognition systems, hearing aids, teleconferencing, and hands-free communication. 

Recent advancements have witnessed the rise of deep learning techniques for multi-channel speech enhancement. Deep neural networks (DNNs)-based multi-channel speech enhancement methods can roughly be divided into three categories: The first category comprises traditional beamformer modules and uses DNNs to assist estimation of spatial statistics \cite{7471664, wang2018all, 10096213, wang2024attention}. The second category follows the so-called filter-and-sum beamformer methodology, utilize DNNs to estimate a filter/mask for each channel and then sum the filtered/masked signals in individual channels \cite{luo2019fasnet, ren2021causal, 9747528, li2022embedding}. The third category consists of end-to-end multi-channel speech enhancement techniques that omit traditional beamformer modules, but instead use convolutional neural networks, recurrent neural networks, and Transformer or Conformer-based architectures for modeling complex spatial and temporal dependencies in speech signals \cite{liu2020multichannel, lee2023deft, 9944916, quan2024spatialnet, 10387735}. The end-to-end paradigm has shown promising results in modeling complex relationships between multi-channel acoustic features and clean speech, without relying on handcrafted features, e.g., inter-channel time/phase/level difference feature \cite{lee2023deft, 9944916, quan2024spatialnet, 10387735}.

The end-to-end multi-channel speech enhancement methods typically depend on the designation of one microphone signal as the reference for processing. 
However, in many realistic scenarios, such as employing a large distributed array of omnidirectional microphones, a small highly directional array (e.g., head-mounted or binaural microphones), or scenarios involving speaker movement or array motion or rotation, the relative speech signal-to-noise ratio (SNR) or signal-to-distortion ratio (SDR) \cite{vincent2006performance} across microphones tends to vary over time or between recordings.
Therefore, the selection of the reference microphone may affect the quality of the enhanced signal. 
Several studies have investigated the reference channel selection problem in distributed microphone arrays for conventional beamformer-based multi-channel speech enhancement, where speaker's position is relatively static \cite{lawin2012reference, 9272831}. 
The methods include choosing the microphone closest to the target source, using the microphone that has the highest input power or SNR, and selecting a reference channel based on maximizing the output SNR. These works have demonstrated the effectiveness of selecting a proper reference channel. Among these approaches, selection based on the maximization of the output SNR seemed to produce optimal results \cite{lawin2012reference, 9272831}. 

In current end-to-end multi-channel speech enhancement
systems, the choice of the reference channel is typically
fixed, especially for mask-based approaches that designate a noisy microphone signal
as the reference to enhance and use the clean speech signal of the corresponding channel to evaluate \cite{lee2023deft, 9944916, 10387735}. This can not automatically
identify the best reference channel to deal with real-world
applications, such as in realistic acoustical conditions
mentioned earlier.
\begin{figure*}[thbp]
  \centering
  \includegraphics[width=1\linewidth, height=4cm]{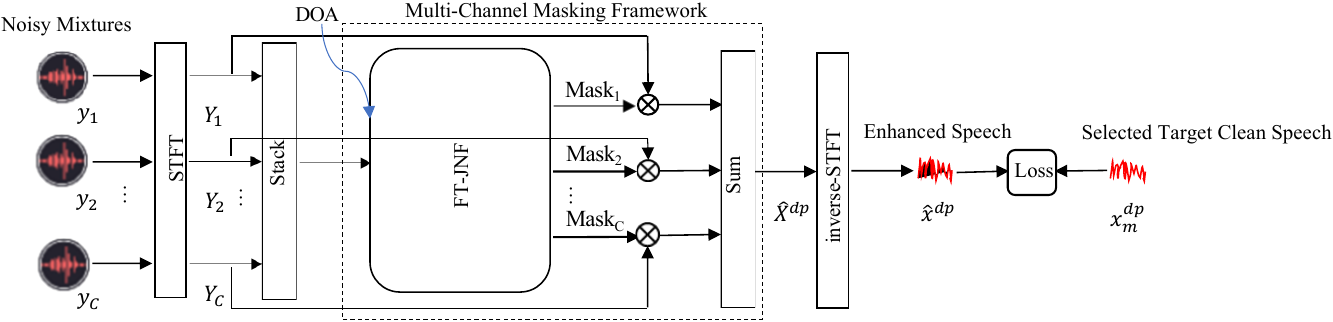}
  \caption{Overview of proposed reference channel selection by multi-channel masking framework for speech enhancement}
\end{figure*}
To address this problem, we propose using the reference channel with the highest output SI-SDR \cite{le2019sdr} during training within a multi-channel masking framework. 
The multi-channel masking strategy draws inspiration from \cite{ren2021causal, 9747528}, in which learning-based deep filter-and-sum beamforming networks predict the complex-valued mask for each input channel, leading to improved performance in comparison to estimating a complex-valued mask for a selected reference channel. 
The dynamic switching of the reference channel for individual audio samples during training allows the model to select/obtain the optimal reference channel during inference. 
Our approach is validated on the simulated dynamic multi-party dialogue D4 dataset of the Spear challenge \cite{9914721}. 
The experimental results on this challenging dataset demonstrate the benefits of our proposed method compared to conventional single-channel masking/filtering method using a fixed reference channel.

The remainder of this paper is organized as follows. Section II defines the problem, signal model, and necessary notations. Section III introduces the proposed reference channel selection with the multi-channel masking method. Section IV presents the experimental setup and results. Section V concludes our work.

\section{Problem Definition and Notations}
We target the reference channel selection problem in the so-called cocktail-party scenario, extracting the speech signal of the target speaker from interfering speech and ambient noise.
The mixture signals are recorded with $C$ microphones. 
We denote the noisy speech signal at the $c$-th ($c=1,.., C$) microphone by ${y}_{c}(t)$, where $t$ is the time index.
The acoustic signal at the $c$-th microphone is given as
\begin{equation}
{y}_{c}(t) = {x}^{dp}_{c}(t) + {v}_{c}(t),
\end{equation}
where ${x}^{dp}_{c}(t)$ is the direct-path target speech signal, ${v}_{c}(t)$ represents the sum of all the early reflections and the late reverberation and all interfering speech signals and noise.
In this work, we assume that the target source can move, so that the acoustic transfer function of the target speaker with respect to the microphone array is time-variant.
Given the noisy recording ${y}_{c}(t)$, we aim to recover the direct-path target speech signal ${x}^{dp}_{c}(t)$.
We represent signals using their complex-valued short-time Fourier transforms (STFTs) in each frequency $f$ and time frame $i$ as:
\begin{equation}
{Y}_{c}(f,i) = {X}^{dp}_{c}(f,i) + {V}_{c}(f,i).
\end{equation}

In this section, we also formulate some necessary notations/concepts described in the following sections: 
$\text{in-SI-SDR}_{c} = \text{SI-SDR$\langle {y}_{c}(t), {x}^{dp}_{c}(t)\rangle$}$ denotes the input (unprocessed) SI-SDR at the $c$-th channel and $\text{out-SI-SDR}_{c} = \text{SI-SDR$\langle \hat{x}^{dp}(t), {x}^{dp}_{c}(t)\rangle$}$ denotes the output (processed) SI-SDR. $\hat{x}^{dp}(t)$ represents the estimated desired signal and can be the estimation of the direct-path speech signal recorded at an arbitrary microphone. SI-SDR$\langle \textit{a}, \textit{b} \rangle$ is the SI-SDR metric calculated using signal $b$ as the reference signal and signal $a$ as the predicted signal.


\section{Proposed Method}
The key idea of the proposed reference channel selection method is to let the network select the channel that obtains the highest out-SI-SDR in the multi-channel masking scheme. 
The overall architecture of the proposed reference channel selection method for multi-channel speech enhancement is depicted in Fig. 1. Note that we omit the time index $t$ and time-frequency indices $(f, i)$ for simplicity in this figure. $Y_{c}$ is the stack of real and imaginary parts of $Y_{c}(f,i)$. 
$Y_{1}$, $Y_{2}$, ..., and $Y_{c}$ are stacked into a vector for each time frame to pass 
the multi-channel masking framework to output a single-channel enhanced speech signal. 

The multi-channel masking framework can be used with any deep neural network that enables estimating a complex mask for each channel.
We choose the FT-JNF network in \cite{10321676} as the mask estimation network. In \cite{10321676}, FT-JNF also utilizes the target speaker's DOA information to assist target speech enhancement. The detailed setup is described in Section IV. 
As seen in Fig. 1, the FT-JNF network estimates the mask for individual channels. After obtaining multiple masks, each estimated complex mask is complex multiplied with the STFT of the corresponding channel. Then the masked STFT of all the channels are summed to produce the enhanced spectra. The enhanced complex spectra are transformed into the final waveform through inverse STFTs.

In the training stage, the ground truth target clean speech signal for calculating loss with an output enhanced speech signal for each audio sample is selected 
based on the channel that has the highest output SI-SDR, i.e., $m$ = ${argmax}_{m} $out-SI-SDR$\langle \hat x^{dp}$, ${x}^{dp}_{m} \rangle$, where $m$ is the channel that produces highest out-SI-SDR.
Specifically, before updating parameters through gradient descent, the model computes the SI-SDR metric between the single-channel output signal and clean reference speech signals of all the channels. The clean reference signal to yield the maximum SI-SDR value is then selected for loss calculation and subsequent weight updates. This process occurs during both forward propagation and backpropagation processing of the network for each audio sample, which is similar to utterance-level permutation invariant training (uPIT) \cite{kolbaek2017multitalker}. 
During the inference stage, based on this training scheme, the network is expected to obtain the optimal reference channel that has the highest out-SI-SDR.
We refer to this method as MM-Auto-out.

\section{Experimental Setup}

\subsection{Dataset}
The experiments are conducted on the simulated dataset D4 of the Spear challenge \cite{9914721}. This dataset features multiple distinct conversations that occur simultaneously, making it more challenging to follow a particular speaker. The dataset uses clean speech from SLR83 corpus, synthesized head movements, and competing dialogues that significantly overlap with the target speech \cite{9914721}. The head-mounted microphones in this dataset have four microphones on a pair of glasses and two additional microphones positioned at the ears. Given that there are only anechoic target speaker ear signal references (direct-path speech signals) as training targets, and considering the signals in the two microphones positioned at the ears have bigger spatial diversity than the other four microphones, we evaluate our proposed method only using the binaural microphones. The original Spear challenge dataset D4 has training, development, and evaluation splits. The clean references of the evaluation set are withheld by the challenge organizers \cite{10248181}. Therefore, the development set is used as the evaluation set. For the training set, there are 9 sessions, each session has 28, 29, or 30 one-minute recordings. For the development set, there are 3 sessions, each session has 28, 29, or 30 one-minute recordings. The dataset also includes the target speaker’s DOA with respect to the array as azimuth and elevation angles sampled every 50 ms.

\subsection{Data Preprocessing}
We first downsample the original one-minute recordings from 48 kHz to 16 kHz. Then, we select segments of the mixture signal with at least one active speaker in conversation according to the participants' VAD labels provided by the dataset. These segments are further divided into 3-second clips for both the training and development sets. The training set has 6824 clips, while the development set has 2136 clips. 
For computing the STFT, we use 512-sample (32 ms) window with a 50\% overlap and the Hann window. The target speaker's DOA is transformed to cartesian coordinates $(x,y,z)$ for each time frame. However, since the DOAs are captured at a 50 ms interval in the dataset, and the frame length is set to 32 ms, we perform linear interpolation on the DOA coordinates $(x,y,z)$.

\subsection{Baseline and Compared Methods}

In this paper, we use the terms SM and MM to refer to single-channel masking and multi-channel masking, respectively. As the dataset utilized in our study features two microphone channels, the best reference channel for each audio sample is either the left or right channel. 
We compare the proposed approach with the following methods: 

(i) SM-Left/Right is the baseline that uses multi-channel inputs to estimate a single-channel complex mask for either the left or right channel as the reference channel.

(ii) MM-Left/Right serves as the ablation experiment of our proposed MM-Auto-out method. It estimates the complex mask for each input channel and then sums all the masked signals but 
fixes the reference channel to be either the left or right.

(iii) SM-Fixed (oracle) can be thought of as an oracle performance benchmark for the SM-Fixed approach on this dataset, where the channel with the highest in-SI-SDR is selected as the reference channel for each audio sample during both training and inference, and the selected reference channel is always put at the first position when input to the network. 

(iv) MM-Auto-in aims to assess the impact of using the highest in-SI-SDR versus out-SI-SDR for reference channel selection. Here, the reference channel is chosen based on the highest in-SI-SDR.

In the evaluation, in our proposed MM-Auto-out method, the clean reference speech signal used for evaluating is chosen following the same rule as in training.
MM-Auto-in follows the same evaluation rule as MM-Auto-out, for its reference channel is not fixed during training.
For SM-Left/Right and MM-Left/Right methods, since the reference channel is fixed, the clean reference speech extracted from the chosen reference channel is employed for assessment.

\subsection{Training and Network Configurations}
The loss function to be minimized is the negative SI-SDR. The Adam optimizer is used to train the models with a learning rate initialized to ${1 \times 10 ^ {-4}}$ and then exponentially decays by 0.99 for each training epoch. A total of 45 epochs with a mini-batch size of one 3-second audio sample is used.

The FT-JNF network serves as the base network for estimating single-channel or multi-channel masks. It consists of two long short-term memory (LSTM) layers operating along the frequency bins and time axis, respectively. The first layer employs bidirectional LSTM, while the second layer uses unidirectional LSTM for efficient online processing.
Input to the FT-JNF network comprises stacked real and imaginary parts of the STFTs from two microphone channels. In addition, a conditioning mechanism similar to \cite{10321676} is employed to leverage the target speaker's DOA information. The DOA $(x,y,z)$ for each time frame is encoded through a linear layer to match the number of units in the F-LSTM layer (set to 256). This encoded DOA information initializes the forward and backward initial states of the bidirectional F-LSTM layer.

For single-channel mask prediction, the network employs a tanh activation function in the final layer to estimate a complex mask, which is then multiplied with the reference channel's noisy STFT to obtain the target speech STFT coefficients. In the case of multi-channel mask estimation, the network similarly uses a tanh activation function to estimate complex masks for each channel. 
Note that the encoded DOA mechanism is applied to our proposed method and all other methods.

\subsection{Results and Discussion}
In order to investigate the benefit of the proposed method in comparison to a fixed reference channel in different ranges of absolute difference of input (unprocessed) SDRs in the left and right microphones,
we define the absolute difference of input SDRs as $\text{in-SDR-Gap}_{j,k} = \lvert \text{in-SDR}_{j} - \text{in-SDR}_{k}\rvert$.
Here, $j$ and $k$ is the left or right microphone channel, and $| \cdot |$ represents absolute value. $\text{in-SDR}_{c} = \text{SDR$\langle {y}_{c}(t), {x}^{dp}_{c}(t)\rangle$}$ denotes input SDR at the $c$-th channel ($c$=$j$ or $c$=$k$).
SDR$\langle \textit{a}, \textit{b} \rangle$ means the SDR metric is calculated using signal $b$ as the reference signal and signal $a$ as the prediction signal.
To facilitate evaluation, we also define output (processed) SDR as $\text{out-SDR}_{c} = \text{SDR$\langle \hat{x}^{dp}(t), {x}^{dp}_{c}(t)\rangle$}$.
We then compute the distribution of in-SDR-Gap in three ranges of [0,3], (3,6], (6,12] dB for all 3-second audio samples in the training and development set. The biggest in-SDR-Gap 
in this dataset is 12 dB. The training and development set has a similar distribution, with proportions falling into three ranges: 58.1\%, 36.6\%, and 5.3\%; and 55.7\%, 36.5\%, and 7.8\% respectively.
\begin{table}[h]
\centering
\caption{Performance of proposed method and other methods with encoded DOA.}
\begin{tabular}{l r r r | r r r}
\hline
\multirow{2}{*}{Methods} & \multicolumn{3}{c|}{out-SI-SDR (dB)} & \multicolumn{3}{c}{out-SDR (dB)} \\
\cline{2-7}
& [0-3] & (3,6] & (6,12] & [0,3] & (3,6] & (6,12] \\
\hline
SM-Left (baseline) & -2.9 & -1.7 & -0.2 & 1.9 & 3.4 & 5.0 \\
SM-Right (baseline) & -3.0 & -1.9 & -0.7 & 2.1 & 3.4 & 4.8 \\
MM-Left & -2.9 & -1.5 & 0.1 & 2.1 & 3.7 & 5.4 \\
MM-Right & -2.7 & -1.6 & -0.2 & 2.3 & 3.8 & 5.3 \\
SM-Fixed (oracle) & -2.7 & -1.4 & \bfseries 0.5 & 2.0 & 3.8 & \bfseries 5.9 \\
MM-Auto-in & -4.5 & -3.6 & -1.8 & 1.3 & 3.1 & 5.4 \\
MM-Auto-out (prop.) & \bfseries -2.5 & \bfseries -1.3 & -0.1 & \bfseries 2.4 & \bfseries 3.9 & 5.4 \\
\hline
\end{tabular}
\end{table}
To compare the proposed method with others, we computed SDR and SI-SDR metrics on 3-second audio samples from the development dataset. Table I presents the results. 
It is noteworthy that the performance of left and right channels chosen as reference channels differs in both SM and MM methods, which is expected. Our approach exhibits clear benefits compared to SM-Left and SM-Right methods. Ablation experiment results of MM methods with fixed reference channels consistently demonstrate better performance compared to corresponding SM-Left/Right methods.
Our proposed method, MM-Auto-out, surpasses MM-Left and MM-Right in most conditions. This shows the benefit of selecting the reference channel with the maximal output SI-SDR during training. Conversely, the MM-Auto-in method performs the least effectively, suggesting that choosing the channel with the maximal input SI-SDR as the reference channel may not yield the optimal enhanced signal.
The manual SM-Fixed (oracle) method excels particularly in large in-SDR-Gap of (6,12], indicating that a fixed reference channel with significantly high input SDR could possibly lead to a higher output SDR. Notably, our proposed method achieves comparable results with SM-Fixed (oracle) in the in-SDR-Gap range of (3,6] and even outperforms the SM-Fixed (oracle) method within the in-SDR-Gap range of [0,3].

To better understand the performance advantages of our proposed method, we calculated the energy and the relative energy proportions between the two channels of masked STFTs across time-frequency bins in all testing samples for MM-Left, MM-Right, and our MM-Auto-out methods: the most energetically dominant masked channel and the least energetically dominant masked channel. Notice that the two masked channels are chosen separately for each audio sample, and the energy has no unit.
The results are summarized in Table II.
\begin{table}[h]
\centering
\caption{Energy and its proportion of two masked channels in MM-Left/Right and MM-Auto-out methods.}
\begin{tabular}{l l l}
\hline
Methods 
& Most Energetic & Least Energetic \\
\hline
MM-Left & 251597 (72.5\%) & 95149 (27.5\%) \\
MM-Right &  191901 (71.7\%) & 75499 (28.3\%) \\
MM-Auto-out (prop.) & 183256 (71.2\%) & 74223 (28.8\%) \\
\hline
\end{tabular}
\end{table}

Across all three methods, it is notable that the energy in the two masked channels both account for a relatively large proportion.
This indicates that the multi-channel masking scheme does spatial filtering in addition to masking. Interestingly, our proposed MM-Auto-out method exhibits very similar but lower energy to MM-Right in both masked channels, 
this is supported by the fact that most (about 93\%) samples in the evaluation set choose the right channel as the reference channel. In summary, by leveraging the multi-channel masking mechanism, our proposed method enables selecting the more proper reference channel while concurrently applying spatial filtering to enhance the signal.
\begin{table}[h]
\centering
\caption{Performance of reference channel fixed methods evaluating use the best reference.}
\begin{tabular}{l r r r | r r r}
\hline
\multirow{2}{*}{Methods} & \multicolumn{3}{c|}{out-SI-SDR (dB)} & \multicolumn{3}{c}{out-SDR (dB)} \\
\cline{2-7}
& [0,3] & (3,6] & (6,12] & [0,3] & (3,6] & (6,12] \\
\hline
SM-Left (baseline) & -2.8 & -1.6 & -0.2 & 1.9 & 3.3 & 5.0 \\
SM-Right (baseline) & -2.9 & -1.8 & -0.7 & 2.0 & 3.4 & 4.8 \\
MM-Left & -2.8 & -1.4 & 0.1 & 2.1 & 3.7 & 5.3 \\
MM-Right & -2.6 & -1.5 & -0.2 & 2.3 & 3.8 & 5.3 \\
SM-Fixed (oracle) & -2.7 & \bfseries -1.3 & \bfseries 0.5 & 2.0 & 3.8 & \bfseries 5.9 \\
\hline
\end{tabular}
\end{table}

To further ensure a fair comparison, we evaluated the MM-Left/Right and SM-Left/Right methods using the same methodology as the MM-Auto methods, i.e., selecting the channel with the maximum out-SI-SDR value and then computing the final out-SI-SDR and out-SDR metrics using that channel. Results are presented in Table III.
Overall, the performance change is minimal in comparison to the results in Table I. Both SM and MM methods show a slight improvement in SI-SDR metric within the in-SDR-Gap of [0,3] and (3,6], but no improvement in SDR metric. However, MM-Left exhibits a decrease in the in-SDR-Gap of (6,12].
In addition, the SM-Fixed (oracle) approach only demonstrates a marginal improvement of 0.1 dB in the out-SI-SDR metric within the in-SDR-Gap range of (3,6].
To sum up, our proposed method maintains advantages over SM and ablation experiment MM methods with fixed reference channels and is superior to the SM-Fixed (oracle) method in relatively small in-SDR-Gap ranges of [0,3] and (3,6].
\section{Conclusions}
We have presented a reference channel selection approach for end-to-end multi-channel speech enhancement, which maximizes the output SI-SDR based on a multi-channel masking mechanism.
The experimental results indicate the proposed method leads to better performance than the conventional selection of a fixed reference channel with a single-channel masking approach. The findings of our study are expected to provide valuable insights into the development of approaches for optimal reference channel selection to enhance the signal quality of desired speech signals in similar settings.

\bibliographystyle{IEEEtran}
\bibliography{conference}

\begin{thebibliography}{10}
\providecommand{\url}[1]{#1}
\csname url@samestyle\endcsname
\providecommand{\newblock}{\relax}
\providecommand{\bibinfo}[2]{#2}
\providecommand{\BIBentrySTDinterwordspacing}{\spaceskip=0pt\relax}
\providecommand{\BIBentryALTinterwordstretchfactor}{4}
\providecommand{\BIBentryALTinterwordspacing}{\spaceskip=\fontdimen2\font plus
\BIBentryALTinterwordstretchfactor\fontdimen3\font minus \fontdimen4\font\relax}
\providecommand{\BIBforeignlanguage}[2]{{%
\expandafter\ifx\csname l@#1\endcsname\relax
\typeout{** WARNING: IEEEtran.bst: No hyphenation pattern has been}%
\typeout{** loaded for the language `#1'. Using the pattern for}%
\typeout{** the default language instead.}%
\else
\language=\csname l@#1\endcsname
\fi
#2}}
\providecommand{\BIBdecl}{\relax}
\BIBdecl

\bibitem{7471664}
J.~Heymann, L.~Drude, and R.~Haeb-Umbach, ``Neural network based spectral mask estimation for acoustic beamforming,'' in \emph{2016 IEEE International Conference on Acoustics, Speech and Signal Processing (ICASSP)}, 2016, pp. 196--200.

\bibitem{wang2018all}
Z.-Q. Wang and D.~Wang, ``All-neural multi-channel speech enhancement.'' in \emph{Interspeech}, 2018, pp. 3234--3238.

\bibitem{10096213}
C.-H. Lee, C.~Yang, Y.~Shen, and H.~Jin, ``Improved mask-based neural beamforming for multichannel speech enhancement by snapshot matching masking,'' in \emph{IEEE International Conference on Acoustics, Speech and Signal Processing (ICASSP)}, 2023, pp. 1--5.

\bibitem{wang2024attention}
Y.~Wang, A.~Politis, and T.~Virtanen, ``Attention-driven multichannel speech enhancement in moving sound source scenarios,'' in \emph{IEEE International Conference on Acoustics, Speech and Signal Processing (ICASSP)}, 2024, pp. 11\,221--11\,225.

\bibitem{luo2019fasnet}
Y.~Luo, C.~Han, N.~Mesgarani, E.~Ceolini, and S.-C. Liu, ``Fasnet: Low-latency adaptive beamforming for multi-microphone audio processing,'' in \emph{2019 IEEE automatic speech recognition and understanding workshop (ASRU)}, 2019, pp. 260--267.

\bibitem{ren2021causal}
X.~Ren, X.~Zhang, L.~Chen, X.~Zheng, C.~Zhang, L.~Guo, and B.~Yu, ``A causal u-net based neural beamforming network for real-time multi-channel speech enhancement.'' in \emph{Interspeech}, 2021, pp. 1832--1836.

\bibitem{9747528}
M.~M. Halimeh and W.~Kellermann, ``Complex-valued spatial autoencoders for multichannel speech enhancement,'' in \emph{IEEE International Conference on Acoustics, Speech and Signal Processing (ICASSP)}, 2022, pp. 261--265.

\bibitem{li2022embedding}
A.~Li, W.~Liu, C.~Zheng, and X.~Li, ``Embedding and beamforming: All-neural causal beamformer for multichannel speech enhancement,'' in \emph{IEEE International Conference on Acoustics, Speech and Signal Processing (ICASSP)}, 2022, pp. 6487--6491.

\bibitem{liu2020multichannel}
C.-L. Liu, S.-W. Fu, Y.-J. Li, J.-W. Huang, H.-M. Wang, and Y.~Tsao, ``Multichannel speech enhancement by raw waveform-mapping using fully convolutional networks,'' \emph{IEEE/ACM Transactions on Audio, Speech, and Language Processing}, vol.~28, pp. 1888--1900, 2020.

\bibitem{lee2023deft}
D.~Lee and J.-W. Choi, ``{Deft-an}: Dense frequency-time attentive network for multichannel speech enhancement,'' \emph{IEEE Signal Processing Letters}, vol.~30, pp. 155--159, 2023.

\bibitem{9944916}
K.~Tesch and T.~Gerkmann, ``Insights into deep non-linear filters for improved multi-channel speech enhancement,'' \emph{IEEE/ACM Transactions on Audio, Speech, and Language Processing}, vol.~31, pp. 563--575, 2023.

\bibitem{quan2024spatialnet}
C.~Quan and X.~Li, ``Spatialnet: Extensively learning spatial information for multichannel joint speech separation, denoising and dereverberation,'' \emph{IEEE/ACM Transactions on Audio, Speech, and Language Processing}, vol.~32, pp. 1310--1323, 2024.

\bibitem{10387735}
H.~N. Chau, T.~D. Bui, H.~B. Nguyen, T.~T.~H. Duong, and Q.~C. Nguyen, ``A novel approach to multi-channel speech enhancement based on graph neural networks,'' \emph{IEEE/ACM Transactions on Audio, Speech, and Language Processing}, vol.~32, pp. 1133--1144, 2024.

\bibitem{vincent2006performance}
E.~Vincent, R.~Gribonval, and C.~F{\'e}votte, ``Performance measurement in blind audio source separation,'' \emph{IEEE Transactions on Audio, Speech, and Language Processing}, vol.~14, no.~4, pp. 1462--1469, 2006.

\bibitem{lawin2012reference}
T.~C. Lawin-Ore and S.~Doclo, ``Reference microphone selection for mwf-based noise reduction using distributed microphone arrays,'' in \emph{Speech Communication; 10. ITG Symposium}.\hskip 1em plus 0.5em minus 0.4em\relax VDE, 2012, pp. 1--4.

\bibitem{9272831}
J.~Zhang, H.~Chen, L.-R. Dai, and R.~C. Hendriks, ``A study on reference microphone selection for multi-microphone speech enhancement,'' \emph{IEEE/ACM Transactions on Audio, Speech, and Language Processing}, vol.~29, pp. 671--683, 2021.

\bibitem{le2019sdr}
J.~Le~Roux, S.~Wisdom, H.~Erdogan, and J.~R. Hershey, ``{SDR}--half-baked or well done?'' in \emph{ICASSP 2019-2019 IEEE International Conference on Acoustics, Speech and Signal Processing (ICASSP)}, 2019, pp. 626--630.

\bibitem{9914721}
P.~Guiraud, S.~Hafezi, P.~A. Naylor, A.~H. Moore, J.~Donley, V.~Tourbabin, and T.~Lunner, ``An introduction to the speech enhancement for augmented reality (spear) challenge,'' in \emph{2022 International Workshop on Acoustic Signal Enhancement (IWAENC)}, 2022, pp. 1--5.

\bibitem{10321676}
K.~Tesch and T.~Gerkmann, ``Multi-channel speech separation using spatially selective deep non-linear filters,'' \emph{IEEE/ACM Transactions on Audio, Speech, and Language Processing}, vol.~32, pp. 542--553, 2024.

\bibitem{kolbaek2017multitalker}
M.~Kolb{\ae}k, D.~Yu, Z.-H. Tan, and J.~Jensen, ``Multitalker speech separation with utterance-level permutation invariant training of deep recurrent neural networks,'' \emph{IEEE/ACM Transactions on Audio, Speech, and Language Processing}, vol.~25, no.~10, pp. 1901--1913, 2017.

\bibitem{10248181}
B.~Stahl and A.~Sontacchi, ``Multichannel subband-fullband gated convolutional recurrent neural network for direction-based speech enhancement with head-mounted microphone arrays,'' in \emph{2023 IEEE Workshop on Applications of Signal Processing to Audio and Acoustics (WASPAA)}, 2023, pp. 1--5.

\end{thebibliography}


\begin{thebibliography}{00}
\bibitem{b1} G. Eason, B. Noble, and I. N. Sneddon, ``On certain integrals of Lipschitz-Hankel type involving products of Bessel functions,'' Phil. Trans. Roy. Soc. London, vol. A247, pp. 529--551, April 1955.
\bibitem{b2} J. Clerk Maxwell, A Treatise on Electricity and Magnetism, 3rd ed., vol. 2. Oxford: Clarendon, 1892, pp.68--73.
\bibitem{b3} I. S. Jacobs and C. P. Bean, ``Fine particles, thin films and exchange anisotropy,'' in Magnetism, vol. III, G. T. Rado and H. Suhl, Eds. New York: Academic, 1963, pp. 271--350.
\bibitem{b4} K. Elissa, ``Title of paper if known,'' unpublished.
\bibitem{b5} R. Nicole, ``Title of paper with only first word capitalized,'' J. Name Stand. Abbrev., in press.
\bibitem{b6} Y. Yorozu, M. Hirano, K. Oka, and Y. Tagawa, ``Electron spectroscopy studies on magneto-optical media and plastic substrate interface,'' IEEE Transl. J. Magn. Japan, vol. 2, pp. 740--741, August 1987 [Digests 9th Annual Conf. Magnetics Japan, p. 301, 1982].
\bibitem{b7} M. Young, The Technical Writer's Handbook. Mill Valley, CA: University Science, 1989.
\end{thebibliography}

\end{document}